\documentclass[useAMS,usenatbib,usegraphicx]{mn2e}

\newcommand{\etalb}{et al.}
\newcommand{\beq}{\begin{equation}}
\newcommand{\beqa}{\begin{eqnarray}}
\newcommand{\eeq}{\end{equation}}
\newcommand{\eeqa}{\end{eqnarray}}

\newcommand{\bx}{{\bf x}}

\newcommand{\bk}{{\bf k}}
\newcommand{\kB}{{k_{\rm B}}}
\newcommand{\Lya}{Ly$\alpha$~}

\newcommand{\td}{{\tilde{\delta}}}

\title[Probing the Epoch of Early Baryonic Infall Through 21cm Fluctuations]
{Probing the Epoch of Early Baryonic Infall Through 21cm Fluctuations}

\author[R. Barkana and A. Loeb]{R. Barkana$^{1}$ and A. Loeb$^{2}$
\thanks{E-mail:
barkana@wise.tau.ac.il (RB); aloeb@cfa.harvard.edu (AL)}\\
$^{1}$School of Physics and Astronomy, The Raymond and Beverly Sackler
Faculty of Exact Sciences, Tel Aviv University, Tel Aviv 69978,
ISRAEL\\ $^{2}$Astronomy Department, Harvard University, 60 Garden
Street, Cambridge, MA 02138, USA}

\begin{document}

\pagerange{\pageref{firstpage}--\pageref{lastpage}} \pubyear{2005}

\maketitle

\label{firstpage}

\begin{abstract}
After cosmological recombination, the primordial hydrogen gas
decoupled from the cosmic microwave background (CMB) and fell into the
gravitational potential wells of the dark matter. The neutral hydrogen
imprinted acoustic oscillations on the pattern of brightness
fluctuations due to its redshifted 21cm absorption of the CMB. Unlike
CMB temperature fluctuations which probe the power spectrum at cosmic
recombination, we show that observations of the 21cm fluctuations at
$z \sim 20$--200 can measure four separate fluctuation modes (with a
fifth mode requiring very high precision), thus providing a unique
probe of the geometry and composition of the universe.
\end{abstract}


\section{Introduction}

Resonant absorption by neutral hydrogen at its spin-flip 21cm
transition can be used to map its three-dimensional distribution at
early cosmic times \citep{Hogan, Scott, Madau}. The primordial
inhomogeneities in the cosmic gas induced variations in the optical
depth for absorption of the cosmic microwave background (CMB) at the
redshifted 21cm wavelength. Absorption occurs as long as the spin
temperature of hydrogen, $T_s$ (which characterizes the population
ratio of the upper and lower states of the 21cm transition), is lower
than the CMB temperature, $T_\gamma$.  This condition is satisfied in
the redshift interval $20\la z\la 200$ \citep{Loeb04}, before the
first galaxies formed in the universe \citep{BL01}.

The formation of hydrogen through the recombination of protons and
electrons at $z\sim 10^3$ decoupled the cosmic gas from its mechanical
drag on the CMB.  Subsequently, the neutral gas was free to fall into
the gravitational potential wells of the dark matter. The infalling
gas retained some memory of the oscillations imprinted in the power
spectrum of its density fluctuations on the scale traversed by sound
waves in the photon-baryon fluid prior to decoupling. The signatures
of these acoustic oscillations have been detected recently in the CMB
anisotropies from $z\sim 10^3$ \citep{WMAP} and in the large-scale
distribution of massive galaxies at $z\sim 0.35$ \citep{eisen,
surveys2}. Here we calculate their imprint on the 21cm brightness
fluctuations at the intermediate range of $20\la z\la 200$ and propose
it as a probe of the composition and geometry of the universe.
Previous calculations of the fluctuations at these redshifts were
inaccurate since they ignored fluctuations in the sound speed of the
cosmic gas. Furthermore, we show that the 21cm power spectrum as a
function of redshift can be decomposed into a combination of five
fixed fluctuation modes, with coefficients that vary with redshift. We
show that these modes and their coefficients can all be measured
directly and then used to constrain cosmological parameters. Several
groups are currently engaged in the construction of low-frequency
radio experiments that will attempt to detect the diffuse 21cm
radiation from $z\ga 6$ (http://space.mit.edu/eor-workshop/).

\section{Growth of Density Perturbations}

On large scales, the dark matter (dm) and the baryons (b) are affected only
by their combined gravity. The evolution of sub-horizon linear
perturbations is described by two coupled second-order differential
equations \citep{Peebles}: \beqa \ddot{\delta}_{\rm dm} + 2H \dot
{\delta}_{\rm dm} & = & 4 \pi G \bar{\rho}_m \left(f_{\rm b} \delta_{\rm b}
+ f_{\rm dm} \delta_{\rm dm}\right)\ , \nonumber \\ \ddot{\delta}_{\rm b}+
2H \dot {\delta}_{\rm b} & = & 4 \pi G \bar{\rho}_m \left(f_{\rm b}
\delta_{\rm b} + f_{\rm dm} \delta_{\rm dm}\right)\ , \eeqa where
$\delta_{\rm dm}(t)$ and $\delta_{\rm b}(t)$ are the perturbations in the
dark matter and baryons, respectively, the derivatives are with respect to
cosmic time $t$, $H(t)=\dot{a}/a$ is the Hubble constant with
$a=(1+z)^{-1}$, and we assume that the mean mass density $\bar{\rho}_m(t)$
is made up of respective mass fractions $f_{\rm dm}$ and $f_{\rm
b}=1-f_{\rm dm}$. Since these linear equations contain no spatial
gradients, they can be solved spatially for $\delta_{\rm dm}(\bx,t)$ and
$\delta_{\rm b}(\bx,t)$ or in Fourier space for $\td_{\rm dm}(\bk,t)$ and
$\td_{\rm b}(\bk,t)$.

Defining $\delta_{\rm tot}
\equiv f_{\rm b} \delta_{\rm b} + f_{\rm dm} \delta_{\rm dm}$ and
$\delta_{\rm b-} \equiv \delta_{\rm b} - \delta_{\rm tot}$ , we find
\beqa \ddot{\delta}_{\rm tot} + 2H \dot {\delta}_{\rm tot} & = & 4 \pi
G \bar{\rho}_m \delta_{\rm tot}\ , \nonumber \\ \ddot{\delta}_{\rm
b-}+ 2H \dot{\delta}_{\rm b-} & = & 0\ .  \eeqa Each of these
equations has two independent solutions. The equation for $\delta_{\rm
tot}$ has the usual growing and decaying solutions, which we denote
$D_1(t)$ and $D_4(t)$, respectively, while the $\delta_{\rm b-}$
equation has solutions $D_2(t)$ and $D_3(t)$; we number the solutions
in order of declining growth rate (or increasing decay rate). We
assume an Einstein-de Sitter, matter-dominated universe in the
redshift range $z=20$--150, since the radiation contributes less than
a few percent at $z < 150$, while the cosmological constant and the
curvature contribute to the energy density less than a few percent at
$z > 3$. In this regime $a \propto t^{2/3}$ and the solutions are
$D_1(t)=a/a_i$ and $D_4(t)=(a/a_i)^{-3/2}$ for $\delta_{\rm tot}$, and
$D_2(t)=1$ and $D_3(t)=(a/a_i)^{-1/2}$ for $\delta_{\rm b-}$, where we
have normalized each solution to unity at the starting scale factor
$a_i$. Intuitively, since baryons and dark matter both feel the same
gravity, the difference between them changes according to $D_2$ and
$D_3$ which describe the evolution of perturbations in the absence of
gravity. We set the initial redshift $z_i=150$, when the photon
density and temperature perturbations become negligible on sub-horizon
scales and can be neglected \citep{NB05}.

The observable baryon perturbation can then be written as \beq
\td_{\rm b}(\bk,t) =
\td_{\rm b-} + \td_{\rm tot} = \sum_{m=1}^4 \td_{m}(\bk)\, D_m(t)\
, \label{eq:delb} \eeq and similarly for the dark matter perturbation,
\beq \td_{\rm dm}={1\over f_{\rm dm}}\left(\td_{\rm
tot}-f_b\td_b\right)= \sum_{m=1}^4 \td_{m}(\bk)\, C_m(t)\ ,
\label{eq:deldm} 
\eeq 
where $C_i=D_i$ for $i=1,4$ and $C_i=-(f_{\rm b}/f_{\rm dm})D_i$ for
$i=2,3$. We establish the values of $\td_{m}(\bk)$ by inverting the
$4\times 4$ matrix ${\bf A}$ that relates the 4-vector
$(\td_1,\td_2,\td_3,\td_4)$ to the 4-vector that represents the
initial conditions $(\td_{\rm b},\td_{\rm dm},\dot{\td}_{\rm
b},\dot{\td}_{\rm dm})$ at the initial time.
%

\section{Growth of Temperature Perturbations}

We include the (previously ignored) fluctuations in the sound speed
of the cosmic gas caused by Compton heating of the gas, which is due
to scattering of the residual electrons with the CMB photons. The
evolution of the temperature $T$ of a gas element of density $\rho_b$
is given by the first law of thermodynamics: \beq dQ = \frac{3} {2}
\kB dT - \kB T d
\log \rho_{\rm b}\ , \eeq where $dQ$ is the heating rate per
particle. Before the first galaxies formed, \beq \frac{dQ} {d t} = 4
\frac{\sigma_{\rm T}\, c} {m_e} \, \kB (T_\gamma - T) \rho_\gamma
x_e(t) \ , \eeq where $\sigma_T$ is the Thomson cross-section,
$x_e(t)$ is the electron fraction out of the total number density of
gas particles, and $\rho_\gamma$ is the CMB energy density at a
temperature $T_\gamma$. In the redshift range of interest, we assume
that the photon temperature ($T_\gamma = T_\gamma^0/a$) is spatially
uniform, since the high sound speed of the photons (i.e.,
$c/\sqrt{3}$) suppresses fluctuations on the sub-horizon scales that
we consider, and the horizon-scale $\sim 10^{-5}$ fluctuations
imprinted at cosmic recombination are also negligible compared to the
smaller-scale fluctuations in the gas density and
temperature. Fluctuations in the residual electron fraction $x_e(t)$
are even smaller.  Thus,
\beq \frac{dT} {dt} = \frac{2} {3} T
\frac{d \log
\rho_{\rm b}} {dt} + \frac{x_e(t)}{t_\gamma}\, (T_\gamma - T)\,
a^{-4}\ , \eeq where $t_\gamma^{-1} \equiv \bar{\rho}_\gamma^0
({8\sigma_{\rm T}\, c}/{3 m_e}) = 8.55 \times 10^{-13} {\rm yr}^{-1}$.
After cosmic recombination, $x_e(t)$ changes due to the slow
recombination rate of the residual ions: \beq {d x_e(t)\over dt} = -
\alpha_B(T) x_e^2(t) \bar{n}_H (1+y)\ , \eeq where $\alpha_B(T)$ is
the case-B recombination coefficient of hydrogen, $\bar{n}_H$ is the
mean number density of hydrogen at time $t$, and $y=0.079$ is the
helium to hydrogen number density ratio. This yields the evolution of
the mean temperature, ${d \bar{T}}/{dt} = - 2 H
\bar{T} + {x_e(t)}{t_\gamma^{-1}}\, (T_\gamma - \bar{T})\, a^{-4}$. In
prior analyses \citep[e.g.,][]{Peebles, Ma} a spatially uniform speed
of sound was assumed for the gas at each redshift. Note that we refer
to $\delta p/ \delta \rho$ as the square of the sound speed of the
fluid, where $\delta p$ is the pressure perturbation, although we are
analyzing perturbations driven by gravity rather than sound waves
driven by pressure gradients. 

Instead of assuming a uniform sound speed, we find the first-order
perturbation equation, \beq \frac{d \delta_T} {d t} = \frac{2}{3}
\frac{d \delta_b} {dt} - \frac{x_e(t)} {t_\gamma} \frac{T_\gamma}
{\bar{T}} a^{-4} \delta_T\ , \label{eq:order1} \eeq
where we defined the fractional temperature perturbation $\delta_T$.  Like
the density perturbation equations, this equation can be solved separately
at each $\bx$ or at each $\bk$. Furthermore, the solution $\delta_T (t)$ is
a linear functional of $\delta_b(t)$ [for a fixed function $x_e(t)$].
Thus, if we choose an initial time $t_i$ then using Eq.~(\ref{eq:delb}) we
can write the solution in Fourier space as \beq \td_T (\bk,t) =
\sum_{m=1}^4 \td_{m}(\bk)\, D^T_m(t) + \td_T (\bk,t_i)\, D^T_0(t)\ ,
\label{eq:delT} \eeq where $D^T_m(t)$ is the solution of
Eq.~(\ref{eq:order1}) with $\delta_T = 0$ at $t_i$ and with the
perturbation mode $D_m(t)$ substituted for $\delta_b(t)$, while
$D^T_0(t)$ is the solution with no perturbation $\delta_b(t)$ and with
$\delta_T = 1$ at $t_i$. By modifying the CMBFAST code \citep{cmbf},
we numerically solve Eq.~(\ref{eq:order1}) along with the density
perturbation equations for each $\bk$ down to $z_i=150$, and then
match the solution to the form of Eq.~(\ref{eq:delT}). We note that
\citet{Indian} derived a similar equation to Eq.~(\ref{eq:order1}) but
solved it only for a density perturbation that follows the growing
mode $D_1(t)$, neglecting the other density modes and the initial
temperature perturbation.

\begin{figure}
\includegraphics[width=84mm]{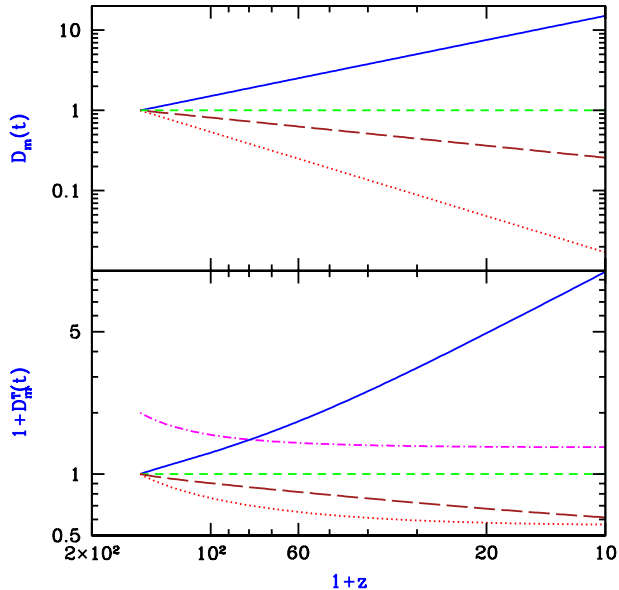}
\caption{Redshift evolution of the amplitudes of the 
independent modes of the density perturbations (upper panel) and the
temperature perturbations (lower panel), starting at redshift 150. We
show $m=1$ (solid curves), $m=2$ (short-dashed curves), $m=3$
(long-dashed curves), $m=4$ (dotted curves), and $m=0$ (dot-dashed
curve). Note that the lower panel shows one plus the mode amplitude.}
\label{fig:Tevol}
\end{figure}

\begin{figure}
\includegraphics[width=84mm]{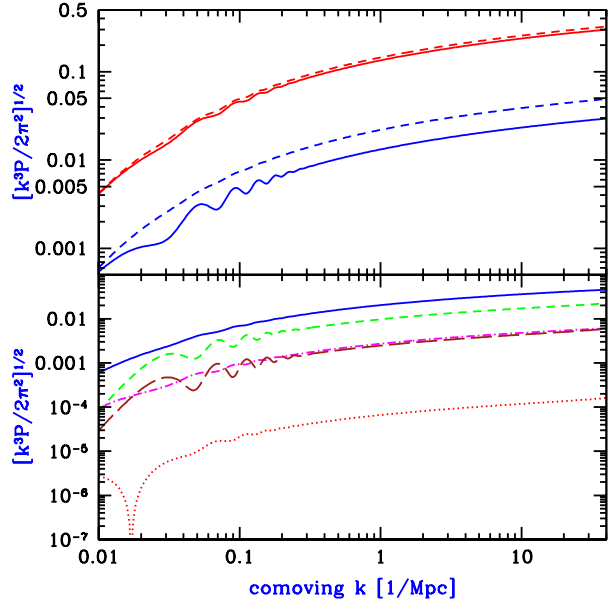}
\caption{Power spectra and initial perturbation amplitudes versus
wavenumber. The upper panel shows $\td_{\rm b}$ (solid curves) and
$\td_{\rm dm}$ (dashed curves) at $z=150$ and 20 (from bottom to top).
The lower panel shows the initial ($z=150$) amplitudes of $\td_1$
(solid curve), $\td_2$ (short-dashed curve), $\td_3$ (long-dashed
curve), $\td_4$ (dotted curve), and $\td_T(t_i)$ (dot-dashed
curve). Note that if $\td_1$ is positive then so are $\td_3$ and
$\td_T(t_i)$, while $\td_2$ is negative at all $k$, and $\td_4$ is
negative at the lowest $k$ but is positive at $k > 0.017$ Mpc$^{-1}$.}
\label{fig:del}
\end{figure}

Figure~\ref{fig:Tevol} shows the time evolution of the various
independent modes that make up the perturbations of density and
temperature, starting at the time $t_i$ corresponding to
$z_i=150$. $D^T_2(t)$ is identically zero since $D_2(t)=1$ is
constant, while $D^T_3(t)$ and $D^T_4(t)$ are
negative. Figure~\ref{fig:del} shows the amplitudes of the various
components of the initial perturbations. We consider comoving
wavevectors $k$ in the range 0.01 -- 40 Mpc$^{-1}$, where the lower
limit is set by considering sub-horizon scales at $z=150$ for which
photon perturbations are negligible compared to $\delta_{\rm dm}$ and
$\delta_{\rm b}$, and the upper limit is set by requiring baryonic
pressure to be negligible compared to gravity. $\td_2$ and $\td_3$
clearly show a strong signature of the large-scale baryonic
oscillations, left over from the era of the photon-baryon fluid before
recombination, while $\td_1$, $\td_4$, and $\td_T$ carry only a weak
sign of the oscillations. For each quantity, the plot shows $[k^3
P(k)/(2 \pi^2)]^{1/2}$, where $P(k)$ is the corresponding power
spectrum of fluctuations.

$\td_4$ is already a very small correction at $z=150$ and declines
quickly at lower redshift, but the other three modes all contribute
significantly to $\td_{\rm b}$, and the $\td_T(t_i)$ term remains
significant in $\td_T(t)$ even at $z \la 100$. Note that at $z=150$
the temperature perturbation $\td_T$ has a different shape with
respect to $k$ than the baryon perturbation $\td_{\rm b}$, showing
that their ratio cannot be described by a scale-independent speed of
sound.

\begin{figure}
\includegraphics[width=84mm]{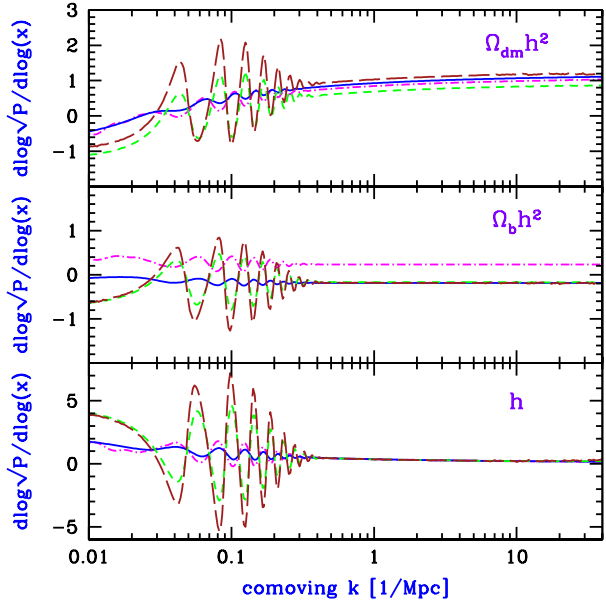}
\caption{Relative sensitivity of perturbation amplitudes at $z=150$ to
cosmological parameters. For variations in a parameter $x$, we show
$d{\rm log}\sqrt{P(k)}/d{\rm log}(x)$. We consider variations in
$\Omega_{\rm dm} h^2$ (upper panel), in $\Omega_b h^2$ (middle panel),
and in the Hubble constant $h$ (lower panel). When we vary each
parameter we fix the other two, and the variations are all carried out
in a flat $\Omega_{\rm total}=1$ universe. We show the sensitivity of
$\td_1$ (solid curves), $\td_2$ (short-dashed curves), $\td_3$
(long-dashed curves), and $\td_T(t_i)$ (dot-dashed curves).}
\label{fig:sens}
\end{figure}

The power spectra of the various perturbation modes and of
$\td_T(t_i)$ depend on the initial power spectrum of density
fluctuations from inflation and on the values of the fundamental
cosmological parameters ($\Omega_{\rm dm}$, $\Omega_b$,
$\Omega_{\Lambda}$, and $h$). If these independent power spectra can
be measured through 21cm fluctuations, this will probe the basic
cosmological parameters through multiple combinations, allowing
consistency checks that can be used to verify the adiabatic nature and
the expected history of the perturbations. Figure~\ref{fig:sens}
illustrates the relative sensitivity of $\sqrt{P(k)}$ to variations in
$\Omega_{\rm dm} h^2$, $\Omega_b h^2$, and $h$, for the quantities
$\td_1$, $\td_2$, $\td_3$, and $\td_T(t_i)$. Not shown is $\td_4$,
which although it is more sensitive (changing by order unity due to
$10\%$ variations in the parameters), its magnitude always remains
much smaller than the other modes, making it much harder to
detect. Note that although the angular scale of the baryon
oscillations constrains also the history of dark energy through the
angular diameter distance, we have focused here on other cosmological
parameters, since the contribution of dark energy relative to matter
becomes negligible at high redshift.

\section{21cm Fluctuations}

The spin temperature $T_s$ is defined through the ratio between the
number densities of hydrogen atoms, ${n_1/ n_0}=(g_1/
g_0)\exp\left\{-{T_\star/ T_s}\right\},$ where subscripts $1$ and $0$
correspond to the excited and ground state levels of the 21cm
transition, $(g_1/g_0)=3$ is the ratio of the spin degeneracy factors
of the levels, and $T_\star=0.0682$K corresponds to the energy
difference between the levels. The 21cm spin temperature is on the one
hand radiatively coupled to the CMB temperature, and on the other hand
coupled to the kinetic gas temperature $T$ through collisions
\citep{AD} or the absorption of \Lya photons \citep{Wout, Field}. For the
concordance set of cosmological parameters \citep{WMAP}, the mean
brightness temperature on the sky at redshift $z$ (relative to the CMB
itself) is \citep{Madau} $T_b = 28\, {\rm mK}\,
[({1+z})/{10}]^{1/2} \left[({T_s - T_{\gamma}})/{T_s}\right]
\bar{x}_{\rm HI}$, where $\bar{x}_{\rm HI}$ is the mean neutral
fraction of hydrogen.

In general, fluctuations in $T_b$ can be sourced by fluctuations in
gas density, temperature, neutral fraction, radial velocity gradient,
and \Lya flux from galaxies. The velocity gradient term
\citep{Sobolev} is
in Fourier space \citep{kaiser, Indian, Indian2} $\td_{d_rv_r} = -\mu^2
{\dot{\td}}/{H}$, where $\mu = \cos\theta_k$ in terms of the angle
$\theta_k$ of $\bk$ with respect to the line of sight.  We can
therefore write the fluctuation as
\citep{BL04} \beq
\label{Tbk} \td_{T_b} (\bk,t) = \mu^2 \dot{\td}(\bk) H^{-1} + 
\beta \td(\bk) + \beta_T \td_T(\bk) + \td_{\rm rad}(\bk)\ , \eeq where we
have defined time-dependent coefficients $\beta$ and $\beta_T$
(combining the relevant explicit terms from Eq.~2 of \citet{BL04}),
and have also combined the terms that depend on the radiation fields
of \Lya photons and ionizing photons into $\td_{\rm rad}$.  Before the
first galaxies formed,
$\bar{x}_{H I}=1$ and $\td_{\rm rad}=0$. Thus, \beqa \td_{T_b} (\bk,t)
& = & \sum_{m=1}^4 \td_{m}(\bk)\, \left[ \mu^2 H^{-1}(t) \dot{D}_m(t)
+ \beta D_m(t) \right. \nonumber \\ & & \left. + \beta_T D^T_m(t)
\right] + \beta_T \td_T (\bk,t_i)\, D^T_0(t)\label{eq:dTb}\ . \eeqa
The power spectrum $P_{T_b}$ is defined by \beq \langle \td_{T_b}
(\bk_1,t) \td_{T_b,t} (\bk_2) \rangle = (2\pi)^3 \delta^D(\bk_1+\bk_2)
P_{T_b}(\bk_1,t)\ . \eeq At $z=150$, the overall contributions of the
$\td_1$, $\td_2$, $\td_3$, and $\td_T (t_i)$ terms to $P_{T_b}$ are
comparable, while the $\td_4$ term (which corresponds to the usual
decaying mode) represents only a $1\%$ correction to the 21cm power
spectrum. The contributions of the different terms to $\td_{T_b}$ then
scale with redshift roughly like the corresponding density modes
$D_m(t)$, with the $\td_T (t_i)$ term varying roughly as $\propto
1/a$, so that the usual growing mode $\td_1$ dominates at low
redshift.

\begin{figure}
\includegraphics[width=84mm]{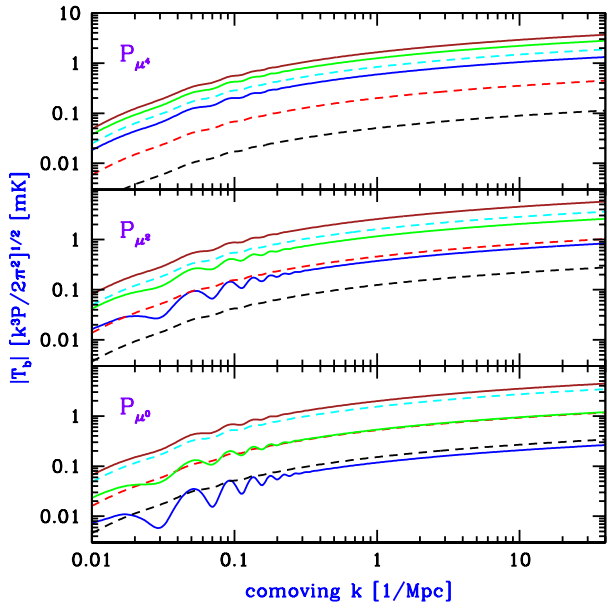}
\caption{Power spectra of 21cm brightness fluctuations versus 
wavenumber. We show the three power spectra that are separately
observable, $P_{\mu^4}$ (upper panel), $P_{\mu^2}$ (middle panel), and
$P_{\mu^0}$ (lower panel). In each case we show redshifts 150, 100, 50
(solid curves, from bottom to top), 35, 25, and 20 (dashed curves,
from top to bottom).}
\label{fig:21cm}
\end{figure}

Noting that observations of the 21cm power spectrum can be analyzed as
a function of $k$ and $\mu$, we write the power spectrum as a
polynomial in $\mu$ \citep{BL04},
\begin{equation}
P_{T_b}(\bk) = \mu^4 P_{\mu^4}(k) + \mu^2 P_{\mu^2}(k) + P_{\mu^0}(k)
\ ,
\end{equation}
and obtain the three separate power spectra of 21cm
fluctuations. Figure~\ref{fig:21cm} shows their values at various
redshifts. The fluctuations rise with time until $z \sim 50$ due to
the growth of fluctuations and then decline at lower redshift due to
the reduced collisional coupling of the 21cm spin temperature to the
gas temperature \citep{Loeb04}. The five power spectra shown in the
lower panel of Fig.~\ref{fig:del} along with the coefficients of the
five modes in Eq.~(\ref{eq:dTb}) can all be determined directly from
observations. Suppose, e.g., that we measure the three observable
power spectra at $N$ $k$-values at $M$ different redshifts, for a
total of $3NM$ data points. As we have shown, these data points can be
modeled in terms of the amplitudes of five separate fluctuation modes
at the same $N$ $k$-values, along with two coefficients per redshift
for each fluctuation mode [corresponding to the contributions to the
$\mu^2$ term and to the $\mu$-independent term in Eq.~(\ref{eq:dTb})].
The total of $5N+10M$ parameters can then be determined as long as the
number of data points is larger than this total. Thus, $M=4$ redshifts
suffice as long as $N \ge 5$, while $M=2$ suffices for any $N \ge 20$.
Measurements at additional redshifts will allow a more accurate
reconstruction along with multiple consistency checks. Note that
Figure~\ref{fig:Tevol} shows that separating out the different modes
at $z \sim 20$ requires an order of magnitude higher measurement
precision than at $z \sim 100$.

Detection of these fluctuations with future experiments \citep{Zalda04,
Miguel1, Miguel2} would set new constraints on the evolution of the
universe through a regime of cosmic time that has never been probed
before. Unlike existing cosmological probes, the three-dimensional
nature of 21cm fluctuations makes possible measurements at a wide
range of redshifts and scales. CMB temperature fluctuations
\citep{WMAP} probe the power spectrum only at cosmic recombination, and
only on scales $k \le 0.2$ Mpc$^{-1}$ which are not damped by photon
diffusion. Galaxy redshift surveys \citep{surveys1, surveys2} probe
the density power spectrum only at low redshifts where the
fastest-growing mode $D_1(t)$ is already strongly dominant, and the
signature of the baryonic oscillations is weak. Furthermore, even on
large scales ($k < 0.1$ Mpc$^{-1}$) the biased distribution of
galaxies may be affected by complex astrophysical feedback processes
such as cosmic reionization \citep{Bar04a}. 

Indeed, CMB fluctuations and galaxy surveys cannot individually
constrain multiple-parameter cosmological models due to degeneracies,
and must be combined in order to achieve significant constraints
\citep{combo}. In contrast, 21cm fluctuations provide a signal at a
sufficiently early epoch when the baryons have not yet caught up with
the dark matter, and four separate fluctuation modes are detectable
(with a fifth mode requiring higher precision). Since each mode
depends differently on the cosmological parameters (see
Fig.~\ref{fig:sens}), such measurements would provide great
discriminatory power in determining the cosmological parameters and in
testing for any deviations from the standard $\Lambda$CDM
paradigm. This cosmological information is in principle available from
measurements in any redshift range, but the amplitudes of some of the
modes decline at low redshift and they become more challenging to
measure (see Fig.~\ref{fig:Tevol}). While we have focused on deriving
constraints from the power spectra, we note that the coefficients in
Eq.~(\ref{eq:dTb}) depend on known cosmological and atomic physics, so
that measuring them from the 21cm data will help to further constrain
the fundamental cosmological parameters.

\section*{Acknowledgments}
This work was supported in part by NSF grants AST-0204514, AST-0071019
and NASA grant NAG 5-13292 . R.B. acknowledges the support of Israel
Science Foundation grant 28/02/01.

\bsp

\label{lastpage}

\end{document}